# The Right to Communications Confidentiality in Europe: Protecting Privacy, Freedom of Expression, and Trust

*Frederik J. Zuiderveen Borgesius\**
*& Wilfred Steenbruggen\*\**

*In the European Union, the General Data Protection Regulation (GDPR) provides comprehensive rules for the processing of personal data. In addition, the EU lawmaker intends to adopt specific rules to protect confidentiality of communications, in a separate ePrivacy Regulation. Some have argued that there is no need for such additional rules for communications confidentiality.*

*This Article discusses the protection of the right to confidentiality of communications in Europe. We look at the right's origins to assess the rationale for protecting it. We also analyze how the right is currently protected under the European Convention on Human Rights and under EU law.*

---

\*   Prof. Dr. Frederik J. Zuiderveen Borgesius is Professor of Law at the Digital Security Group (DiS), at the Radboud University Nijmegen, and researcher at the Institute for Information Law (IViR), at the University of Amsterdam. We thank our colleagues for their comments on drafts of this article: Lillian Edwards, Irene Kamara, Adam Shinar, Peter Swire, Joris van Hoboken, the editors of Theoretical Inquiries in Law, all participants at the conference '*The Problem of Theorizing Privacy*, 8-9 January 2018, and all the participants at the Privacy Law Scholars Conference Europe, 27 January 2018, Brussels, and at the Privacy Law Scholars Conference, 30-31 May 2018 Washington. We thank Sabine Rijnen and Sonja van Harten for research assistance. Cite as: Frederik J. Zuiderveen Borgesius & Wilfred Steenbruggen, *The Right to Communications Confidentiality in Europe: Protecting Privacy, Freedom of Expression and Trust*, 20 Theoretical Inquiries L. 291 (2019).

\*\*  Dr. Wilfred Steenbruggen is an attorney at Bird & Bird LLP (The Hague, the Netherlands). He wrote his PhD thesis on communications confidentiality. *See, infra* note 11.

*We show that at its core the right to communications confidentiality protects three individual and collective values: privacy, freedom of expression, and trust in communication services. The right aims to ensure that individuals and organizations can safely entrust communication to service providers. Initially, the right protected only postal letters, but it has gradually developed into a strong safeguard for the protection of confidentiality of communications, regardless of the technology used.*

*Hence, the right does not merely serve individual privacy interests, but also other more collective interests that are crucial for the functioning of our information society. We conclude that separate EU rules to protect communications confidentiality, next to the GDPR, are justified and necessary.*

## Introduction

This Article discusses the protection of the right to confidentiality of communications in Europe. In 2017, the European Commission published a proposal for an ePrivacy Regulation, which includes rules to protect the confidentiality of electronic communications.[1] Some have argued that there is no need for an ePrivacy Regulation, because the General Data Protection Regulation (GDPR), is sufficient to protect privacy.[2]

This Article focuses on the following questions: What is the right to communications confidentiality, and what is the right's scope? What is the rationale for the right to communications confidentiality and how is the right protected in current European law? Are separate EU rules, in addition to the GDPR, needed to protect the confidentiality of communications? We focus on electronic communications. National rules are outside the scope of this Article.[3]

---

1   *Proposal for a Regulation of the European Parliament and of the Council Concerning the Respect for Private Life and the Protection of Personal Data in Electronic Communications and Repealing Directive 2002/58/EC (Regulation on Privacy and Electronic Communications)*, COM (2017) 10 final (Jan. 10, 2017), https://ec.europa.eu/digital-single-market/en/news/proposal-regulation-privacy-and-electronic-communications [hereinafter *ePrivacy Proposal* 2017].
2   For instance, 63.4% of industry respondents to the consultation by the European Commission see no need for special rules for the electronic communications sector on confidentiality of electronic communications. See *id.* at ¶ 3.2.
3   On the law on confidentiality in the UK and the U.S., see Daniel J. Solove & Neil M. Richards, *Privacy's Other Path: Recovering the Law of Confidentiality*,

In Part I, we explore the right's origins and the rationale for protecting communications confidentiality. We argue that communications confidentiality is important, not only because it protects privacy but also because it protects other key values for the information society. By ensuring that individuals and businesses can freely exchange information and ideas with others, the right protects certain aspects of freedom of expression. Moreover, the right aims to ensure that information of any nature can be safely entrusted to communication service providers. Such trust in communication services is not only important to protect individual rights such as privacy and freedom of expression, but also for the proper functioning of our information society.

Parts II and III turn to current law. For readers who are not familiar with the complicated human rights regime in Europe, we introduce the European Convention on Human Rights (ECHR) and the European Court of Human Rights (ECtHR) in Part II. We also introduce the European Union, its Charter of Fundamental Rights, its Court of Justice of the European Union (CJEU), and its ePrivacy rules.

Part III analyzes the scope of the right to communications confidentiality under the European Convention on Human Rights and under EU law. We discuss the following aspects: (A) To which technologies does the right apply? (B) Do communications have to be private to qualify for protection? (C) To what extent does the right to communications confidentiality protect metadata? (D) Does the right still apply after the transmission has ended?

We conclude that separate EU rules to protect the right to communications confidentiality are not only justified, but also necessary.

## I. Foundations of the Right to Communications Confidentiality

### A. Historical Roots of the Right to Communications Confidentiality

Almost all European constitutions contain a right protecting the confidentiality of communications.[4] As a constitutional right, communications confidentiality is connected to the former state postal monopoly. The history of the post is

---

   96 Geo. Wash. L.J. 123 (2007).

4  *See, e.g.*, 1831 Const. art. 22 (Belg.); Gw. [Constitution] art. 13 (Neth.); GG [Constitution] art. 10 (Ger.), https://www.gesetze-im-internet.de/gg/GG.pdf; art. 15 Costituzione [Cost.] (It.). On communications confidentiality in national constitutions, see also Bert-Jaap Koops et al., *A Typology of Privacy*, 38 U. Pa. J. Int'l L. 483 (2016). *See European Data Protection Supervisor Opinion on the Proposal on Privacy and Electronic Communications (ePrivacy Regulations)*, similar in the whole of Europe.[5] In the Middle Ages, the right to provide postal services was seen as a regal privilege, like the right of coinage. As with other privileges, the commercial exploitation of the postal right was first entrusted to private parties, such as the *House of Thurn und Taxis* which built an extensive postal network, covering large parts of Europe.[6]

However, kings and princes realized that the postal service was a lucrative source of income and started to nationalize it. King Louis XI of France was the first to do so (in 1464).[7] In other countries this example was followed. This was the beginning of a long tradition in which the provision of postal services, and later also telecommunications services, was seen as an exclusive government task.

The post was already safeguarded by special guarantees with regard to confidentiality and security before it became an exclusive government task.[8] This protection was not so much established for the protection of secrets of the heart revealed in letters, but rather for the benefit of trust in trade.[9] The rise of the post coincides with the rise of trade and trade fairs in the early Middle Ages. Tradesmen needed a periodical provision of information about what was happening in the important European trading centers. The post constituted the main infrastructure for the exchange of trade information.

The postal monopoly enabled the state to intercept letters. Indeed, before the French Revolution, authorities in Europe had habitually read the letters that were entrusted to the state postal service.[10] Given the significance of the post as infrastructure for the provision of information, this can be regarded as a counterpart to the censorship of print that was common in Europe at

---

　　　6/2017 (Apr. 24, 2017), https://edps.europa.eu/sites/edp/files/publication/17-04-24_eprivacy_en.pdf [hereinafter *European Data Protection Supervisor Opinion*].

5　　*See* Johannes Hofman, Vertrouwelijke Communicatie: Een Rechtsvergelijkende Studie Over De Geheimhouding Van Communicatie In Grondrechtelijk Perspectief Naar Internationaal, Nederlands En Duits Recht [Confidential Communication: A Comparative Study on the Confidentiality of Communications in Constitutional Perspective in International, Dutch and German Law] 13-45 (1995).

6　　*Id.*

7　　*Id.*

8　　Paulus van der Velden, Het geheim der brieven Aan de geschiedenis en aan de beginselen van het Staats- en Strafregt getoetst [The Secret of Letters Tested Against the History and the Principles of Constitutional and Penal Law] 102 (1859).

9　　*See* Hofman, *supra* note 5, at 23.

10　*See* David Kahn, The Codebreakers: The Comprehensive History of Secret Communication from Ancient Times to the Internet (1996).

that time, with which the absolutist rulers of the time tried to prevent the distribution of unwelcome information and opinions.[11] The fact that the state monopolized the post was decisive for the recognition of communications confidentiality as a constitutional right.

In response to the large-scale government interception of letters, a right to confidentiality of correspondence was first included in a number of preliminary drafts of the French *Declaration des Droits de l'Homme et du Citoyen de 1789*.[12] The drafters of the *Declaration* did not see the right to confidentiality of correspondence as a privacy-related right. As Ruiz notes, "[w]hen the secrecy of telecommunications first arose as a constitutional right it was not openly regarded as an aspect of the right to privacy, but as an aspect of the freedom of opinion and of expression."[13] The right was "regarded as an aspect of individual freedom and, in particular, as inherent to the freedom to express one's opinions via the post."[14] The final version of the *Declaration,* however, did not include a right to confidentiality of correspondence. According to Ruiz, a separate confidentiality right "was finally considered redundant and thus excluded from the final version of the *Declaration* of 1789: this right was seen implicitly in the recognition in art. 11 of the freedom to express one's opinion in writing."[15]

---

11  For a historical overview of censorship, see Egbert Dommering, *Grensoverschrijdende censuur: het EHRM en oude en nieuwe media* [*Crossborder Censorship: The ECtHR and Old and New Media*], 2013 Auteurs & Média 177; Jan Corbet Et Al., Censures: actes du colloque du 16 mai 2003 [Censorship: Report of the Conference of 16 May 2003] (2003). On the history of the legal protection of confidentiality of communications, see Wilfred Steenbruggen, Publieke Dimensies Van Privé-Communicatie: Een Onderzoek Naar De Verantwoordelijkheid Van De Overheid Bij De Bescherming Van Vertrouwelijke Communicatie In Het Digitale Tijdperk (2009) [Public Dimensions of Private Communication: An Investigation Into the Responsibility of the Government in the Protection of Confidential Communications in the Digital Age], https://pure.uva.nl/ws/files/752230/70945_proefschrift.pdf. Hofman, *supra* note 5, at 23; Blanca Ruiz, Privacy in Telecommunications: A European and an American Approach 64-70 (1997); Axel Arnbak, Securing private communications: protecting private communications security In Eu law — fundamental rights, functional value chains And market incentives (2015), http://dare.uva.nl/search?metis.record.id=492674.

12  Assemblée constituante, Déclaration des droits de l'homme et du citoyen du de 1789 [Declaration of the Rights of Man and of the Citizen of 1789].

13  *See* Ruiz, *supra* note 11, at 68.

14  *Id.* at 67.

15  *Id.* at 68.

The nineteenth century was the era of the great constitutional codifications in Europe. These constitutions aimed at restricting state power and guaranteeing civil rights against the state. The right to confidentiality of correspondence has been included in most European constitutions from that time as one of those classic civil rights that require the state to refrain from intervening in the spheres of individuals.[16] Since then, most constitutions have not changed much, probably because this is difficult to do and often requires a qualified majority in and several readings by Parliament. But in some European countries, the scope of the constitutional right to confidentiality of communications was extended to include the new communication technologies of telephone and telegraph, which were also brought under the state monopoly.[17]

After World War II, international human right treaties introduced a new level of protection of communications confidentiality. In these treaties, communications confidentiality was framed as a privacy-related right. The Universal Declaration of Human Rights from 1948 mentioned privacy, family, home and correspondence in one provision.[18] A similar right to privacy was codified in several human rights treaties such as the 1966 International Covenant on Civil and Political Rights[19] and the 1950 European Convention on Human Rights.[20]

Since the 1970s, another field of law, related to privacy, has become increasingly important: data protection law (comparable with what U.S. scholars might call information privacy law).[21] In the early 1970s, several European countries adopted data protection laws.[22] The EU sees a task for itself in this field, and in 1995 the EU adopted the Data Protection Directive, which contains rules for the processing of personal data.[23] The GDPR replaced that

---

16  *See* HOFMAN, *supra* note 5, at 13-45.
17  In the Netherlands, for example, the right to secrecy in telephone and telegraph was codified in Article 13 of the Dutch Constitution in 1983. *See* STEENBRUGGEN, *supra* note 11, at 242.
18  G.A. Res. 217 (III) A, Universal Declaration of Human Rights, art. 12 (Dec. 10, 1948).
19  International Covenant on Civil and Political Rights, art. 17, Dec. 16, 1966, 999 U.N.T.S. 171.
20  European Convention for the Protection of Human Rights and Fundamental Freedoms, art. 8, Nov. 4, 1950, E.T.S. No. 5. [Hereinafter ECHR].
21  DANIEL J. SOLOVE & PAUL SCHWARTZ, INFORMATION PRIVACY LAW (2017).
22  On the history of the right to protection of personal data, see GLORIA GONZÁLEZ FUSTER, THE EMERGENCE OF PERSONAL DATA PROTECTION AS A FUNDAMENTAL RIGHT OF THE EU (2014).
23  Directive 95/46/EC of the European Parliament and of the Council of 24 October 1995 on the Protection of Individuals with Regard to the Processing of Personal directive and aims at more harmonization and a stronger protection against unfair processing of personal data.[24]

The EU not only aims to protect personal data; it also aims to protect the confidentiality of communications. In the 1990s, the EU liberalized the telecommunications markets. EU law required member states to phase out the state monopoly and to allow free market forces in the telecommunications sector.[25] As a result, member states lost a large part of their previous powers to regulate these markets. The EU stepped in with rules that aim at effective competition and at safeguarding interests that cannot be left to free market forces, such as communications confidentiality and consumer protection.[26]

In 1997, the EU adopted the ISDN Directive,[27] which was replaced in 2002 by the ePrivacy Directive (last updated in 2009).[28] The ePrivacy Directive provides, as part of the EU regulatory framework for the electronic communications sector, specific rules for confidentiality of communications and for other privacy-related subjects in the electronic communications sector. The ePrivacy Regulation, proposed in 2017, aims to further harmonize the rules regarding communications confidentiality and to broaden the scope of these rules to new communication technologies.[29]

To sum up: historically, the right to communications confidentiality is tied to the former state monopoly on postal and telecommunications services. At first, only postal letters were protected, but as telecommunications were brought under the scope of the state monopoly, the scope of the right to communications confidentiality was extended too. (Part III discusses the scope of the right in more detail.) Nowadays, the right is not connected with the state monopoly anymore, but is based on other rationales. The next section discusses why even after the liberalization of telecommunications services it is still important to protect communications confidentiality.

---

    Data and on the Free Movement of Such Data, 1995 O.J. (L 281) 31 (EU) [hereinafter Directive 95/46/EC].

24  Council Regulation 2016/679 2016 O.J. (L 119) 1 (EU) [hereinafter GDPR].

25  *See* Steenbruggen, *supra* note 11, at 167.

26  *See* Christian Koenig & Andreas Bartosch, EC Competition And Telecommunications Law (2002).

27  Directive 97/66/EC of the European Parliament and of the Council of 15 December 1997 Concerning the Processing of Personal Data and the Protection of Privacy in the Telecommunications Sector, 1997 O.J. (L 24) 1.

28  Directive 2002/58/EC of the European Parliament and of the Council of 12 July 2002 Concerning the Processing of Personal Data and the Protection of Privacy in the Electronic Communications Sector, 2002 O.J. (L 201), revised by Directive 2009/136/EC, 2009 O.J. (L 337) 11 [hereinafter Directive 2002/58/EC].

29  *See ePrivacy Proposal* 2017, *supra* note 1.

## B.	Rationales: Privacy, Freedom of Expression, and Trust

Looking at the origins of the right of communications confidentiality, we can distinguish three rationales, which remain valid even after the liberalization of the postal and telecommunication markets. These rationales are: (1) privacy, (2) freedom of expression, and (3) trust in communication services.

*1. Privacy*

Nowadays, privacy is clearly an important rationale for protecting confidentiality of communications. Monitoring, listening, or reading people's communications will often involve infringing their privacy. Electronic communications are an important means of expressing private thoughts and feelings and developing relationships with others. The right to communications confidentiality provides a barrier against unwanted access to these private thoughts and feelings, and thereby protects individual privacy.[30]

*2. Freedom of Expression*

Freedom of expression is another important rationale for protecting the right to communications confidentiality. Freedom of expression includes the "freedom to hold opinions and to receive and impart information and ideas."[31] The ECtHR and the CJEU have confirmed that communications confidentiality is important for freedom of expression.[32]

However, the relationship between freedom of expression and communications confidentiality is complicated. On the one hand, the right to communications confidentiality can be seen as an important facilitator of freedom of expression.[33] The right aims to guarantee, for example, that people can freely exchange

---

30	*See* STEENBRUGGEN, *supra* note 11, at 45.
31	Charter of Fundamental Rights of the European Union, art. 11, Dec. 7, 2000, 55 O.J. 391 [hereinafter EU Charter]. *See id.* at 46-49.
32	*See infra* Part III.
33	*See also* Frank La Rue, Human Rights Council, Report of the Special Rapporteur on the Promotion and Protection of the Right to Freedom of Opinion and Expression, 20, A/HRC/23/40 (Apr. 17, 2013), www.ohchr.org/Documents/ HRBodies/HRCouncil/RegularSession/Session23/A.HRC.23.40_EN.pdf. On the connection between communications confidentiality and freedom of expression, see also STEENBRUGGEN, *supra* note 11, at 46-49; Joris van Hoboken & Frederik Zuiderveen Borgesius, *Scoping Electronic Communication Privacy Rules: Data, Services and Values*, 6 J. INTELL. PROP. INFO. TECH. & ELECTRONIC COM. L. 198 (2015). For a U.S. perspective on the connection between confidentiality, privacy, and freedom of expression, see NEIL RICHARDS, INTELLECTUAL PRIVACY: RETHINKING CIVIL LIBERTIES IN THE DIGITAL AGE 136-152 (2015).

politically sensitive information without fearing interception and prosecution by the authorities.

On the other hand, the right to communications confidentiality may conflict with freedom of expression. A conflict could arise, for instance, if a journalist wanted to access telephone conversations between U.S. President Trump and Prince Mohammed bin Salman of Saudi Arabia regarding the disappearance of journalist Jamal Khashoggi. In such cases, careful consideration must be given to whether a conflict actually exists and, if this is indeed the case, it must be determined which interest weighs more heavily in the case at hand. (Similar conflicts can arise between privacy and freedom of expression).

For this Article, however, we focus on the role of communications confidentiality as a facilitator of freedom of expression.

*3. Trust in Communication Services*

Next to privacy and freedom of expression, trust in communication services has also always been an important rationale for the protection of communications confidentiality. As Arnbak notes, "[t]he underlying rationale to warrant such protection is that communicants entrust communication to an intermediary, thus losing control in relation to the intermediary or third parties."[34] People need to be able to trust that their communication is safe in the hands of a service provider. That was important under the state monopoly, but it is even more so in our interconnected society, which depends heavily on electronic communication services for communications, commercial transactions, e-government, and democratic participation.

If people cannot reasonably assume that electronic communication services are safe, they are likely to be less willing to use these services. Such a chilling effect on communication would threaten not only privacy and freedom of expression, but also other rights and freedoms, such as freedom of thought and freedom of assembly. Those rights and freedoms would be threatened if people's fear of eavesdropping or surveillance led to their not feeling free to discuss their thoughts or their assemblies.

These three rationales — privacy, freedom of expression, and trust — also underlie the protection of communications confidentiality in the ePrivacy Regulation. The ePrivacy Regulation is one of the instruments of the European Commission's Digital Single Market strategy, which aims to "to increase trust in . . . . digital services."[35] The Commission may have thought mainly of the economic dimension of trust. The Parliament, however, stressed that communications confidentiality is important for various other rights and interests, and proposed to add to the preamble of the ePrivacy regulation: "The protection of confidentiality of communications is an essential condition for the respect of other connected fundamental rights and freedoms, such as the protection of freedom of thought, conscience and religion, freedom of assembly, freedom of expression and information."[36]

---

34   Arnbak, *supra* note 11, at 226. See also Steenbruggen, *supra* note 11, at 354.
35   *See ePrivacy Proposal* 2017, *supra* note 1, at ¶ 1.1 (explanatory memorandum to the proposal for an ePrivacy Regulation).

In sum, protection of communications confidentiality is justified and necessary to safeguard not only individual rights to privacy and freedom of expression, but also the public interest of trust in communication services.

## II. General Framework on Communications Confidentiality in Europe

### A. The European Convention on Human Rights

This Part introduces the general framework for the protection of the right to communications confidentiality in Europe, starting with the European Convention on Human Rights (ECHR). In Part III we will further discuss the scope of the right to communications confidentiality under ECHR and EU law.

The ECHR was adopted in 1950 by the Council of Europe, an intergovernmental organization founded by a number of Western European states after World War II, with the aim of international cooperation. Inspired by the Universal Declaration of Human Rights, the Council of Europe drafted the ECHR, which codified a series of freedoms and rights, such as freedom of expression, freedom of association, and the prohibition of torture. Now, the Council of Europe has 47 member states.[37] The ECHR includes a strong enforcement structure: after having exhausted national legal remedies, citizens in treaty states have the right to file a complaint with the ECtHR, whose rulings are binding on the treaty states. Therefore, the ECHR has become an important factor in the protection of fundamental rights in Europe.[38]

---

36  Marju Lauristin (Special Rapporteur), Comm. On Civil Liberties, Justice and Home Affairs, *Report on the Proposal for a Regulation of the European Parliament and of the Council Concerning the Respect for Private Life and the Protection of Personal Data in Electronic Communications and Repealing Directive 2002/58/EC (Regulation on Privacy and Electronic Communications)*, COM (2017) 0010 (Oct. 10, 2017), http://www.europarl.europa.eu/sides/getDoc.do?pubRef=-//EP//NONSGML+REPORT+A8-2017-0324+0+DOC+PDF+V0//EN.

37  *See Who We Are*, Council Europe, https://www.coe.int/en/web/about-us/who-we-are (last visited Mar. 26, 2018).

38  For an extensive introduction into the ECHR, see, for example, Theory and Practice of the European Convention on Human Rights (Pieter van Dijk, Fried

Confidentiality of communications is one of the rights guaranteed in the ECHR. This right is codified in Article 8 of the ECHR, which grants people the right to respect for their private and family life, home, and their *correspondence*:

> 1. Everyone has the right to respect for his private and family life, his home and his correspondence.
>
> 2. There shall be no interference by a public authority with the exercise of this right except such as is in accordance with the law and is necessary in a democratic society in the interests of national security, public safety or the economic well-being of the country, for the prevention of disorder or crime, for the protection of health or morals, or for the protection of the rights and freedoms of others.

Before the 1970s, almost all complaints regarding violation of Article 8 of the ECHR were declared unfounded. Later, the ECtHR gradually expanded the protection offered by Article 8. First, the ECtHR has stated that the ECHR provisions must be "interpreted and applied in a manner which renders its rights practical and effective, not theoretical and illusory."[39] The meaning of the ECHR rights is therefore not fixed. According to the ECtHR, "the Convention is a living instrument which must be interpreted in the light of present-day conditions."[40] The ECtHR says it takes "a pragmatic, common-sense approach rather than a formalistic or purely legal one."[41]

This dynamic interpretation of the ECtHR has increased the scope of the rights in the ECHR, including the right to privacy in Article 8.[42] The ECtHR has brought all kinds of claims under the protection of Article 8 of the ECHR, such as claims regarding formal recognition of a gender change and claims for protection against noise or environmental pollution.[43]

---

    van Hoof, Arjen van Rijn & Leo Zwaak eds., 5th ed. 2018).
39  Christine Goodwin v. United Kingdom, 2002-VI Eur. Ct. H.R. at ¶ 74; *See also* Soering v. United Kingdom, 161 Eur. Ct. H.R. (ser. A) (1989); Armonas v. Lithuania, 2008 Eur. Ct. H.R. 1526, ¶ 38.
40  Matthews v. United Kingdom, 1999-I Eur. Ct. H.R. at ¶ 39. The Court started the "living instrument" approach in Tyrer v. United Kingdom, 26 Eur. Ct. H.R. (ser. A) at ¶ 31 (1978).
41  Botta v. Italy, 1998-I Eur. Ct. H.R. at ¶ 27.
42  In this Article, we use "privacy" and "private life" interchangeably. *See* GONZÁLEZ FUSTER, *supra* note 22, at 82-84, 255 (on the slight difference between "private life" and "privacy.").
43  *See Goodwin*, 2002-VI Eur. Ct. H.R. at ¶ 74; Amann v. Switzerland, 2000-II Eur. Ct. H.R. at ¶ 65.

Furthermore, the ECtHR has developed a strict test for limitations on the rights of Article 8. The test consists of three steps, which are based on the criteria mentioned in the second paragraph of Article 8.[44] In a typical case, somebody complains that his or her privacy is limited by a state to such an extent that the right is violated. In step 1, the ECtHR checks whether the limitation of the right to privacy is prescribed by a sufficiently precise law. The Court also looks whether there are sufficient safeguards to prevent abuse and arbitrariness. In step 2, the ECtHR assesses whether the limitation is necessary in a democratic society. The ECtHR checks whether the limitation serves a pressing social need, and whether principles of proportionality and subsidiarity are met. In step 3, the ECtHR checks whether the limitation serves one of the legitimate aims (from the second paragraph of Article 8 of the ECHR), such as the protection of national security.[45]

Sometimes the state must take action to comply with the ECHR. The ECtHR has derived from the ECHR not only negative obligations (to refrain from action) for the government, but also positive obligations (to take action). As the Court puts it:

> Although the object of Article 8 is essentially that of protecting the individual against arbitrary interference by the public authorities, it does not merely compel the State to abstain from such interference; in addition to this primarily negative undertaking, there may be positive obligations inherent in an effective respect for private and family life.[46]

States may fail to meet their obligations under Article 8 of the ECHR if they do not take sufficient action to protect citizens against infringements by other citizens. As the ECtHR puts it, "there may be positive obligations inherent in an effective respect for private life. These obligations may involve the adoption of measures designed to secure respect for private life even in the sphere of the relations of individuals between themselves . . . ."[47]

---

44 The Court also applies those criteria when dealing with other human rights, in particular those in articles 8-11 of the ECHR.
45 *See, e.g.*, Handyside v. United Kingdom, 24 Eur. Ct. H.R. (ser. A) (1976); Klass v. Germany, 28 Eur. Ct. H.R. (ser. A) (1978); Sunday Times v. United Kingdom, 30 Eur. Ct. H.R. (ser. A) (1979); Silver and others v. United Kingdom, 61 Eur. Ct. H.R. (ser. A) (1983); Malone v. United Kingdom, 95 Eur. Ct. H.R. (ser. A) (1984).
46 *See, e.g.*, X&Y v. Netherlands, 91 Eur. Ct. H.R. (ser. A) at ¶ 23 (1985).
47 Bărbulescu v Romania, CE:ECHR:2016:0112JUD006149608 (referred to Grand Chamber on 6 June 2016). *See also* Z v. Finland, 1997-I Eur. Ct. H.R. at ¶ 36 (1997); Mosley v. United Kingdom, 2011 Eur. Ct. H.R.774, ¶ 106 (2011).

Such measures might, for example, involve adopting appropriate laws and enforcing them.[48]

These positive obligations can lead to a "horizontal" effect of the rights in Article 8 of the ECHR.[49] A private party cannot sue another private party under the ECHR.[50] But citizens can complain to the ECtHR if the state fails to protect their rights against infringements by other citizens (or companies).[51] This way, fundamental rights also regulate the horizontal relationship between individuals.

The positive obligations also apply to the right to communications confidentiality under Article 8 of the ECHR.[52] Treaty states must therefore actively protect this right. In consideration of the economic and technological developments, the right to communications confidentiality can only offer effective protection if the state actively takes measures to protect the right, also in the sphere of individuals between themselves. Effective protection of communications confidentiality is only possible if the law limits what communication providers can do with our communications and requires such providers to protect our communications against unlawful access by third parties.

There is much debate on the question whether human rights (should) have horizontal effect. Some fear that such horizontal application of human rights leads to a "devaluation" of these rights.[53] For others, it is "self-evident" that human rights have horizontal effect.⁵⁴ We agree that in modern society, active government measures are often indispensable to enable people to enjoy their human rights. Regardless of what scholars say on the topic, the ECtHR gives human rights a certain horizontal effect via its case-law on positive obligations.

---

48  *See, e.g.*, *X&Y*, 91 Eur. Ct. H.R. (ser. A) (1985).
49  Jean-François Akandji-Kombe, Positive Obligations Under the European Convention on Human Rights (2007); Paul De Hert, From the Principle of Accountability to System Responsibility? Key Concepts in Data Protection Law and Human Rights Law Discussions (Jan. 8, 2010) (unpublished manuscript), http://www.vub.ac.be/LSTS/pub/Dehert/410.pdf.
50  ECHR, *supra* note 20, at art. 34.
51  The Court says it "does not consider it desirable, let alone necessary, to elaborate a general theory concerning the extent to which the Convention guarantees should be extended to relations between private individuals *inter se*." VGT Verein Gegen Tierfabriken v. Switzerland, VI Eur. Ct. H.R. at ¶ 46 (2001).
52  Positive obligations regarding communications confidentiality can already be derived from Golder v. United Kingdom, Eur. Ct. H.R. (ser. A) (1975). In this case, the Court ruled that the state should take active measures to enable prisoners to correspond. *See also* Cotlet v. Romania, CE:ECHR:2003:0603JUD003856597; Craxi v. Italy (No. 2), CE:ECHR:2003:0717JUD002533794. On positive obligations and the right of communications confidentiality, see, extensively, Steenbruggen, *supra* note 11.
53  *See, e.g.*, Bart Jan de Vos, Horizontale werking van grondrechten: een kritiek [Horizontal Effects of Human Rights: A Critique] 289-97 (2010).

Because the ECtHR interprets and applies Article 8 of the ECHR generously, this provision plays an important role in the protection of the right to privacy and communications confidentiality in Europe. We will discuss to a greater extent what the scope of protection is in Part III. First, we will introduce the other major source of protection of communication confidentiality in Europe.

### B. EU Law

The European Union has 28 member states.⁵⁵ The origins of the EU lie in the postwar situation after World War II when several European countries started the European Coal and Steel Community (1951), the European Economic Community (1957) and the European Atomic Energy Community (1957). These treaties primarily aimed at economic cooperation but impacted many other policy areas as well. The name of the European Communities was changed to European Union in 1993. Each of the EU member states is also a member state of the Council of Europe, and party to the ECHR.⁵⁶

The Court of Justice of the European Union (CJEU) is based in Luxembourg.⁵⁷ The CJEU has the final say on the interpretation of EU law to ensure it is applied in the same way in all EU member states. National judges in the EU can, and in some cases must, ask the CJEU for advice on how to interpret EU law.⁵⁸

---

54  Serge Gutwirth, Privacy and the Information Age 38 (Raf Casert trans., 2001).
55  Soon it is expected to change to 27 member states as the United Kingdom intends to leave the EU (Brexit).
56  Our Member States, Council Europe, http://www.coe.int/en/web/about-us/our-member states (last visited Mar. 27, 2018).
57  Consolidated Version of the Treaty on European Union, art. 13(1), June 6, 2016, 2016 O.J. (C 202), 13 [hereinafter TEU].
58  Consolidated version of the Treaty on the Functioning of the European Union, art. 267, June 7, 2016, 2016 O.J. (C 202) 47 [hereinafter TFEU] (establishing the preliminary reference procedure that differentiates between the right and the duty of national courts to seek a preliminary ruling. Under the discretionary reference stipulated in article 267(2), a national "court or tribunal" may ask the CJEU to give a preliminary ruling if it considers that a decision on the question is "necessary" to enable it to give a judgment in a particular case. The obligatory reference (duty to refer) is established in two cases: with respect to national courts adjudicating at last instance (Article 267(3)) and with respect to all courts faced with a question of the validity of EU law. The obligation of national

As said, the original EU treaties focused mostly on economic issues. However, the CJEU introduced fundamental rights into EU law judgments by accepting them as general principles of EU law in its case-law. The CJEU found these principles in the common constitutional traditions of the member states and the ECHR.[59] The EU is not a party to the ECHR, and therefore the CJEU is strictly speaking not bound to follow the interpretation of the ECtHR. However, in practice, both courts try to prevent conflicts and diverging interpretations, and regularly cite each other's case law.[60]

The EU developed its own fundamental rights document: the Charter of Fundamental Rights of the European Union, which was adopted in 2000 and has been legally binding since 2009.[61] The EU Charter contains all of the rights of the ECHR and a number of other fundamental rights. (We use the phrases "fundamental rights" and "human rights" interchangeably in this Article; the difference is minimal.[62]) EU institutions must comply with the EU Charter. Member states must also comply with the EU Charter, if they act within the scope of EU law (for example, when they are implementing and applying EU directives or regulations).[63]

The EU Charter contains two provisions which protect privacy-related interests. Article 7 of the EU Charter contains a right to privacy, which copies, almost verbatim, Article 8 of the ECHR. The EU Charter uses the more modern and technology-neutral term "communications" instead of "correspondence" (the word used in the ECHR). Article 7 of the EU Charter reads as follows: "Everyone has the right to respect for his or her private and family life, home and communications."

---

    courts of last instance to refer for a preliminary ruling when a question of the interpretation of EU law arises is subject to certain exceptions. *See also* Agne Limante, *Recent Developments in the Acte Clair Case Law of the EU Court of Justice: Towards a More Flexible Approach*, 54 J. Common Mkt. Stud. 1384 (2016).

59  Case T-11/70, Internationale Handelsgesellschaft mbH v. Einfuhr und Vorratsstelle für Getreide und Futtermittel, 1970 E.C.R. 1125, ¶ 4; Case T-4/73, Nold KG v. Commission No. 2, 1974 E.C.R. 491.

60  Since 2009, an explicit basis was included in the TEU to enable the EU to join the ECHR. However, the CJEU ruled in 2014 that the draft accession agreement was not compatible with EU law. See opinion 2/2013, EU:C:2014:2454. It does not seem likely that the EU will join the ECHR anytime soon.

61  *See* TEU, *supra* note 57, at art. 6.1.

62  On the difference between "human rights" and "fundamental rights," see González Fuster, *supra* note 22, at 82-84, 164-66.

63  Aklagaren v. Akerberg Fransson, EU:C:2013:105.

The EU Charter contains a separate provision that lists the criteria for restrictions on its rights.[64] These criteria resemble the criteria for restrictions in the second paragraph of Article 8 of the ECHR.[65] The EU Charter states that, insofar as the Charter contains fundamental rights which correspond with rights in the ECHR, the meaning and scope of these rights are the same as those of the ECHR.[66] However, the EU Charter allows the EU to provide more protection than the ECHR.[67]

In addition to this right to privacy, the EU Charter contains a separate fundamental right to personal data protection. Article 8 of the EU Charter reads as follows:

> 1. Everyone has the right to the protection of personal data concerning him or her.
>
> 2. Such data must be processed fairly for specified purposes and on the basis of the consent of the person concerned or some other legitimate basis laid down by law. Everyone has the right of access to data which has been collected concerning him or her, and the right to have it rectified.
>
> 3. Compliance with these rules shall be subject to control by an independent authority.

The EU has a long and strong tradition of protecting personal data. The 1995 Data Protection Directive[68] has set the bar for the protection of personal data worldwide.[69] The General Data Protection Regulation[70] replaced the Data Protection Directive and the various implementation laws in the member states in May 2018. The GDPR contains rules regarding the processing of personal data, to give effect to Article 8 of the EU Charter (the right to protection of personal data). The GDPR does not aim to protect the right to privacy in general, or to protect communications confidentiality (Article 7 of the EU Charter).

---

64 EU Charter, *supra* note 31, at art. 52. *See also*, Note from the Praesidium, comments on Article 7 (Praesidium 2000) (Explanations Relating to the Complete Text of the Charter), http://www.europarl.europa.eu/charter/pdf/04473_en.pdf.
65 On the difference between article 52 of the EU Charter and article 8(2) of the ECHR, see González Fuster, *supra* note 22, at 201.
66 EU Charter, *supra* note 31, at art. 52(3).
67 *Id.*
68 Directive 95/46/EC, *supra* note 23.
69 *See* Michael D. Birnhack, *The EU Data Protection Directive: An Engine of a Global Regime*, 24 Computer L. & Security Rev. 508 (2008); Michael Birnhack, *Reverse Engineering Informational Privacy Law*, 15 Yale J.L. & Tech. (2012); Anu Bradford, *The Brussels Effect*, 107 Nw. U. L. Rev. 1 (2012).
70 GDPR, *supra* note 24.

Since 1997, the EU has set specific rules for the protection of privacy, communications confidentiality, and personal data in the electronic communications sector. These rules are currently laid down in the ePrivacy Directive (last revised in 2009).[71] The ePrivacy Directive's rules complement and particularize the general rules on personal data protection.[72] The ePrivacy Directive's central provision for the protection of communications confidentiality is Article 5(1). In short, Article 5(1) says that member states must guarantee, through national law, the confidentiality of communication (and related metadata) by means of public electronic communications networks and services. Article 5(1) thus contains a positive obligation, a duty of care, for member states to protect communications confidentiality. In particular, they must prohibit listening, tapping, storage and other kinds of interception or surveillance of communications and the related metadata. The provision thus emphasizes member states' positive obligations regarding confidentiality of communications.[73]

The principle of communications confidentiality is further protected by other provisions of the ePrivacy Directive, which lay down, for instance, strict rules for the processing of the metadata generated by the use of electronic communication services (traffic and location data).[74] In addition, Article 15 contains strict requirements for restrictions by member states for national security, crime prevention, and similar purposes. These requirements resemble the requirements in Article 8(2) of the ECHR. The ePrivacy Directive requires member states to transpose its rules into national law and to apply and enforce this national law.[75]

---

71  Directive 2009/136/EC of the European Parliament and of the Council of 25 November 2009 Amending Directive 2002/22/EC on Universal Service and Users' Rights Relating to Electronic Communications Networks and Services, Directive 2002/58/EC Concerning the Processing of Personal Data and the Protection of Privacy in the Electronic Communications Sector and Regulation (EC) No 2006/2004 on Cooperation Between National Authorities Responsible for the Enforcement of Consumer Protection Laws, 2009 O.J. (L 337) 11.
72  *See* Directive 2002/58/EC, *supra* note 28, at art. 1(2).
73  On positive obligations, §II A of this article.
74  *See id.* at arts. 6, 8-9.
75  In principle, in the spheres of individuals between themselves, the directive will take effect through the national implementation law. Under the case law of the Court of Justice of the European Union, in principle a private person cannot directly invoke a provision of a directive against another private person. Case C-91/92 Faccini Dori v. Recreb Srl., 1994 E.C.R. I-3325.

In 2017, the European Commission published a proposal for a new ePrivacy Regulation, which should replace the ePrivacy Directive.[76] At the time of writing, the ePrivacy proposal is still being discussed in Brussels, and it is unclear when it will be adopted.[77]

### III. THE SCOPE OF THE RIGHT TO COMMUNICATIONS CONFIDENTIALITY

In this Part, we analyze the scope of the right to communications confidentiality. We discuss the scope of the right under Article 8 of the ECHR on the one hand, and under EU law on the other. More specifically, we discuss (A) which technologies are protected; (B) whether communications need to be private in order to be protected; (C) whether the protection includes content only or also the metadata; and (D) whether communication is also protected after transmission.

**A. Which Communication Technologies?**

Originally, the right to communications confidentiality only applied to postal letters. Nowadays, however, people use all kinds of communication technologies and services for personal communication. Are all these communication technologies protected?

*1. Article 8 of the ECHR*
As noted above, Article 8 of the ECHR protects "correspondence." The drafters of the ECHR intended to protect letters.[78] However, by applying a dynamic interpretation, the ECtHR has also brought other communication technologies under the scope of Article 8 of the ECHR.

---

76 *ePrivacy Proposal* 2017, *supra* note 1.
77 On 20 October 2017, the European Parliament adopted its position on the ePrivacy proposal. Generally speaking, the Parliament's version is more privacy-friendly than the Commission's proposal. Marju Lauristin (Special Rapporteur), Comm. On Civil Liberties, Justice and Home Affairs, *Report on the Proposal for a Regulation of the European Parliament and of the Council Concerning the Respect for Private Life and the Protection of Personal Data in Electronic Communications and Repealing Directive 2002/58/EC (Regulation on Privacy and Electronic Communications)*, COM (Oct. 20, 2017), http://www.europarl.europa.eu/sides/getDoc.do?pubRef=-//EP//NONSGML+REPORT+A8-2017-0324+0+DOC+PDF+V0//EN [hereinafter *Parliament Proposal*].
78 COLLECTED EDITION OF THE "TRAVAUX PRÉPARATOIRES" OF THE EUROPEAN CONVENTION ON HUMAN RIGHTS (Council of Eur. ed., 1975-1985).

In 1978, the ECtHR brought phone calls under the scope of Article 8 for the first time. The Court ruled that "although telephone conversations are not expressly mentioned in paragraph 1 of Article 8 . . . . such conversations are covered by the notions of 'private life' and 'correspondence' referred to by this provision."[79] Later, the ECtHR used the same approach to bring telexes,[80] pager messages,[81] and private radio broadcasting[82] within the scope of Article 8 of the ECHR. And in the 2007 *Copland* case the ECtHR considered email and internet use to be covered by the notions of 'private life' and 'correspondence.'[83] In its 2017 *Barbulescu* judgment, the ECtHR has confirmed that Article 8 also protects communications via an instant messaging service.[84]

Hence, the ECtHR has adopted a technology-neutral approach in which there is ample room to protect various means of communication under Article 8 of the ECHR. Indeed, the ECtHR has stated that "Article 8 ECHR protects the confidentiality of private communications, *whatever the content of the correspondence concerned, and whatever form it may take*. This means that what Article 8 protects is the confidentiality of all the exchanges in which individuals may engage for the purposes of communication."[85] In sum, all kinds of communication technologies are protected by Article 8 of the ECHR.

---

79  Klass v. Germany, 28 Eur. Ct. H.R. (ser. A) at ¶ 41 (1978).
80  Christie v. the United Kingdom, App. No. 21482/93, 78 Eur. Comm'n H.R. Dec. & Rep. 119 (1994). This was not a judgment by the ECtHR, but a decision of the European Commission of Human Rights. Until 1998, the admissibility of a complaint was first assessed by the European Commission on Human Rights. Only if the Commission concluded that the complaint was sufficiently well-founded would the Court deal with it. In 1998, the proceedings before the Commission were abolished (Protocol 11 to the ECHR).
81  Taylor-Sabori v. the United Kingdom, 2002 Eur. Ct. H.R. 691.
82  X and Y v. Belgium, App. No. 8962/80, 28 Eur. Comm'n H.R. Dec. & Rep. 112 (1982).
83  Copland v. United Kingdom, 2007-I Eur. Ct. H.R. at ¶ 41:
    According to the Court's case law, telephone calls from business premises are prima facie covered by the notions of "private life" and "correspondence" for the purposes of Article 8 . . . . It follows logically that e-mails sent from work should be similarly protected under Article 8, as should information derived from the monitoring of personal internet usage.
84  Barbulescu v. Romania, CE:ECHR:2017:0905JUD006149608, ¶ 74. This is a Grand Chamber judgment. Judgments by the Grand Chamber have more weight than judgments of other chambers of the Court. In exceptional situations, the Grand Chamber of the Court decides on cases. ECHR, *supra* note 20, at arts. 30, 43.

*2. EU Law*

Article 7 of the Charter of Fundamental Rights of the European Union essentially copies Article 8 of the ECHR, while replacing the word "correspondence" with "communications." The EU lawmaker introduced this change in view of technological developments and did not intend a material change as compared to Article 8 of the ECHR. As the EU Charter aims to give at least as much protection as the ECHR,[86] it can safely be assumed that Article 7 of the EU Charter protects all kinds of personal communication, regardless of the technology used.

The current EU ePrivacy rules, however, have a narrower scope. Many rules in the ePrivacy Directive apply only to providers of publicly available electronic communications networks and electronic communications services. This also goes for Article 5(1), the main provision for communications confidentiality in the ePrivacy Directive[87] and the rules for processing metadata. An electronic communications service is, roughly summarized, a service that consists wholly or mainly of the conveyance of signals on electronic communications networks.[88] In practice, mostly phone companies and internet access providers fall within the scope of "electronic communications services."

Webmail services (e.g., Gmail), instant messaging services (e.g., WhatsApp), and computer-based Voice over IP services (e.g., Skype) are often called "over the top" services. Such over the top services are generally assumed to fall outside the scope of "electronic communications service" in the ePrivacy Directive. Over the top services do not consist of conveying signals on an electronic communication network, but are considered to be independent services delivered on top of the electronic communication network — users can access over the top services via the networks provided by their electronic communication service providers.[89]

---

85   Michaud v. France, CE:ECHR:2012:1206JUD001232311, ¶ 90 (internal citations omitted; emphasis added). *See also* M.N. and others v. San Marino, E:ECHR:2015:0707JUD002800512, ¶ 51-55.
86   *See* EU Charter, *supra* note 31, at art. 52.
87   There is a discussion on the scope of Article 5(1). *See* Steenbruggen, *supra* note 11, at 179; Frederik Zuiderveen Borgesius, Improving Privacy Protection in the Area of Behavioural Targeting (2015), at 175; Arnbak, *supra* note 11.
88   Directive 2002/58/EC, *supra* note 28, at art. 2(c) (referring to the definition in Directive 2002/21/EC of the European Parliament and of the Council of 7 March 2002 on a Common Regulatory Framework for Electronic Communications Networks and Services, 2002 O.J. (L 108) 33).

While the concept of "electronic communications service" is defined narrowly, the EU legislator nevertheless tried to provide reasonably technology-neutral protection to communications confidentiality. Article 5(1) of the ePrivacy Directive requires member states "to ensure the confidentiality of communications and the related traffic data [metadata] by means of a public communications network and publicly available electronic communications services, through national legislation." The term "communication" is defined in the ePrivacy Directive as "any information exchanged or conveyed between a finite number of parties by means of a publicly available electronic communications service."[90]

Under the ePrivacy Directive, communications confidentiality would arguably also apply to forms of communication which traditionally would not be seen as personal communications, but rather as mass communications. For instance, the definition of communication in the ePrivacy Directive includes information by means of a broadcasting service (e.g., TV content) if this information can be related to an identifiable subscriber or user receiving the information.[91] This means that IP-based TV services could also fall within the scope of Article 5 of the ePrivacy Directive (and Article 6 which sets strict rules for the processing of metadata).[92] This shows that the ePrivacy Directive grants a rather broad protection to communications confidentiality.

The 2017 proposal for an ePrivacy Regulation aims to broaden the scope of the right to communications confidentiality even further to all current and future means of communication, regardless of the technology used. The preamble states: "The principle of confidentiality should apply to current and future means of communication, This would include calls, internet access,

---

[89] This may change. For instance, a German judge asked the CJEU for a preliminary ruling on the question whether a webmail service such as Gmail is an "electronic communication service." *EuGH soll Pflichten von Web mail-Anbietern klären*, Jutsiz-online (Feb. 26 2018), http://www.ovg.nrw.de/behoerde/presse/pressemitteilungen/05_180226/index.php (Gr.). On over the top services, see Ilsa Godlovitch et al., Over-the-top (OTTs) Players: Market Dynamics And Policy Challenges (2015), http://www.europarl.europa.eu/RegData/etudes/STUD/2015/569979/IPOL_STU(2015)569979_EN.pdf.

[90] *See* Directive 2002/58/EC, *supra* note 28, at art. 2(d). All kinds of personal communication by means of a publicly available electronic communications service fall within the scope of "communication," regardless of the type of network used (telecommunications, broadcasting, satellite, etc.).

[91] *See id.* (the second part of the definition of communication).

[92] *See* Steenbruggen, *supra* note 11, at 181, 354.

instant messaging applications, e-mail, internet phone calls and personal messaging provided through social media."[93] For that purpose, the notion of "electronic communication service" "which is the central notion determining the scope of the EU electronic communications framework, is significantly broadened and also includes interpersonal communications services, such as Voice over IP, messaging services, and web-based e-mail services.[94]"

Pursuant to the ePrivacy proposal, the right to communications confidentiality would even apply to machine-to-machine communications and many Internet of Things scenarios.[95] This illustrates that the European Commission sees communications confidentiality as more than just a privacy right. The right to communications confidentiality also aims to promote trust in communications. The proposal's preamble states:

> Connected devices and machines increasingly communicate with each other by using electronic communications networks (Internet of Things) . . . . In order to ensure full protection of the rights to privacy and confidentiality of communications, and to promote a trusted and secure Internet of Things in the digital single market, it is necessary to clarify that this Regulation should apply to the transmission of machine-to-machine communications. Therefore, the principle of confidentiality enshrined in this Regulation should also apply to the transmission of machine-to-machine communications.[96]

The European Parliament, however, proposed to delete the recital about machine-to-machine communications, without offering a clear explanation. It remains to be seen to what extent the right to communications confidentiality will apply to machine-to-machine communications in the final version of the ePrivacy Regulation. While there is still much discussion on what the scope should be, so far, the different EU institutions agree that the ePrivacy Regulation should have a wider scope than the ePrivacy Directive.[97]

---

93	*See ePrivacy Proposal* 2017, *supra* note 1, at recital 1.
94	*ePrivacy Proposal* 2017, *supra* note 1, at art. 3(aa). On "electronic communication services," see Frederik Zuiderveen Borgesius et al., An Assessment of the Commission's Proposal on Privacy and Electronic Communications 35-42 (2017). http://www.europarl.europa.eu/RegData/etudes/STUD/2017/583152/IPOL_STU(2017)583152_EN.pdf
95	*See ePrivacy Proposal* 2017, *supra* note 1, at recital 12.
96	*See id.* at recital 12.
97	The definition of "communication" does not return in the ePrivacy Regulation. And the proposal no longer explicitly states that broadcasting services are within its scope if the information can be related to an identifiable user. Here, at first glance Directive 2002/58/EC seems to be offering more protection. However,

In the light of the rationales of the right of communications confidentiality, it makes sense to protect confidentiality of communications, regardless of the technology.[98] Extending the protection of communications confidentiality to new communication services is an essential step towards effective protection.

## B. Only Private Communication?

Communication services are used not only for private communications, but also for professional business purposes or to spread public information and opinions. We have argued that the right to communications confidentiality protects not only privacy, but also freedom of expression and, more generally, trust in communication services. This raises the question: does communication need to be private, or privacy-related, to be protected?

### 1. Article 8 of the ECHR

Case law of the ECtHR shows that the protection of communications confidentiality under Article 8 of the ECHR extends beyond the private sphere. In its *Niemietz* case of 1992, the ECtHR brought the seizure of business letters by the police under the protection of Article 8. The ECtHR noted that Article 8 "does not use, as it does for the word 'life,' any adjective to qualify the word 'correspondence.'"[99] Correspondence is thus protected, regardless of whether it is private. Business emails are also protected, as illustrated by the 2017 *Barbulescu* case.[100]

---

considering that the ePrivacy Regulation proposal generally wants to continue or extend the protection of communications confidentiality as compared to Directive 2002/58/EC, we assume that such broadcasting services would also be covered under the proposal.

98  Obviously, the ePrivacy rules have a narrower scope than Article 8 of the ECHR and Article 7 of the EU Charter, which both apply to all means of personal communications, regardless of the technology used. The ePrivacy Directive and the proposed ePrivacy Regulation focus on electronic communications; the rules do not apply to letters, for instance.

99  Niemietz v. Germany, 251-B Eur. Ct. H.R. (ser. A) ¶ 32 (1992) (internal citations omitted).

100  Barbulescu v. Romania, CE:ECHR:2017:0905JUD006149608, ¶ 72:

> In a number of cases relating to correspondence with a lawyer, [the Court] has not even envisaged the possibility that Article 8 might be inapplicable on the ground that the correspondence was of a professional nature. Furthermore, it has held that telephone conversations are covered by the notions of "private life" and "correspondence" within the meaning of Article 8. In principle, this is also true where telephone calls are made from or received on business premises. The same applies to emails sent from the workplace,

Next to that, the ECtHR has indicated that surveillance of communications threatens not only privacy-related interests, but also freedom of expression: the "menace of surveillance can be claimed in itself to restrict free communication."[101] This illustrates that protecting communications confidentiality is about more than only protecting privacy-interests. In sum: communication does not need to be private to be protected under Article 8 of the ECHR, which is triggered by the mere use of a communication technology, regardless of the nature of the information that is exchanged.

*2. EU Law*

Article 7 of the EU Charter offers at least the same protection as Article 8 of the ECHR. The CJEU says that the right to privacy under EU law also extends to professional activities, similar to the right in the ECHR.[102]

The ePrivacy Directive also protects *any* information exchanged — private or not. Moreover, the ePrivacy Directive aims to protect the interests of both natural and legal persons.[103] The proposed ePrivacy Regulation takes the same approach.[104] In this regard, the ePrivacy rules go beyond the scope of the GDPR. The GDPR only applies to "personal data," defined as "any information relating to an identified or identifiable natural person (…)."[105] In principle, the GDPR thus does not protect data regarding legal persons.[106]

---

which enjoy similar protection under Article 8, as does information derived from the monitoring of a person's internet use (internal citations omitted).

101 *See* Klass v. Germany, 28 Eur. Ct. H.R. (ser. A) at ¶ 37 (1978). *See also* Big Brother Watch & Others v. UK, App. No. 58170/13, 62322/14 and 24960/1, ECLI:CE:ECHR:2018:0913JUD005817013, http://hudoc.echr.coe.int/eng?i=001-186048.

102 For example, the CJEU noted that "the need for protection against arbitrary or disproportionate intervention by public authorities in the sphere of the private activities of any person, whether natural or legal, constitutes a general principle of Community law." And "for the purposes of determining of the scope of this principle regard must be had to the case law of the Court of Human Rights." Case C-94/00, Roquette Frères SA v. Directeur General de la Concurrence, 2002 E.C.R. I-9011

103 Directive 2002/58/EC aims at protecting the fundamental rights of natural persons as users of public electronic communication services, as well as legal persons as subscribers to public electronic communication services. Directive 2002/58/EC, *supra* note 28, at recital 12. *See also ePrivacy Proposal* 2017, *supra* note 1, at recital 7.

104 *See, e.g., id.* at art. 1(1).

105 GDPR, *supra* note 24, at art. 4(1).

106 *Id.* at recital 14. However, sometimes data about a legal person can be data about an individual (personal data) too, for instance when one person runs a company.

Confidentiality of communications is crucial for companies — but the GDPR might not offer them protection. For example, if a company sends an electronic message to buy stocks on the stock market, and no personal data are involved in that communication, the communication could be outside the GDPR's scope. But the company has an interest in keeping that communication confidential. The ePrivacy rules protect such interests, because they also protect business communication.

In conclusion, under both the ECHR and EU law, communication is protected, regardless of its nature. It makes sense that the right to communications confidentiality protects more than only private communications. If only private communication were protected, it would be necessary to access the communication to determine whether or not it deserves protection. Then, the content of communication would already be disclosed, thereby infringing privacy and confidentiality. Considering the values behind communications confidentiality — privacy, freedom of expression, and trust in communication services — the law should indeed protect all communication technologies, regardless of the nature of the information which is exchanged.

### C. Content Only or Metadata Too?

Should the right to communications confidentiality protect only content, or also metadata? Originally, the state postal company was not allowed to open letters — but had to read the address data to deliver the letter. The content of the communication therefore seems more sensitive than the metadata. However, in the context of electronic communications, metadata can be exceedingly sensitive. Nowadays, service providers process a lot of metadata to provide their services. These metadata can be sensitive, are easily processed and analyzed, and may give detailed insights into people's communications, interests and behavior. Does the right of communications confidentiality protect metadata?

#### 1. Article 8 of the ECHR

Metadata are protected by the ECtHR under Article 8 of the ECHR. The 1984 *Malone* case concerned "metering" records, metadata about phone calls. The ECtHR ruled that Article 8 of the ECHR protects such metadata: "the records of metering contain information, in particular the numbers dialled, which is an integral element in the communications made by telephone."[107]

In the *Malone* case, the ECtHR distinguished the processing of metadata (metering records) from intercepting communication, because the telephone operator may legitimately obtain metadata, for instance to ensure that the subscriber is correctly charged. As a result, the ECtHR subjected the interference with metadata to a less strict test. The ECtHR used a similar approach in the case *P.G & J.H.* of 2002.[108]

---

[107] *See* Malone v. United Kingdom, 95 Eur. Ct. H.R. (ser. A) (1984).

*Malone* and *P.G & J.H.* concerned phone records, but in the *Copland* case of 2007, the ECtHR confirmed that Article 8 of the ECHR also protects metadata regarding email and internet usage.[109] It is generally assumed that all metadata will be protected under Article 8 of the ECHR, regardless of the communication technology used.

In sum, metadata are protected under art. 8 ECHR, but when assessing an interference the ECtHR works from the assumption that capturing communications metadata will normally constitute a less serious infringement than capturing communications content. To some extent, that distinction is understandable, because some metadata need to be processed by the service provider in order to provide the service.

*2. EU Law*

In recent years, the CJEU has delivered important judgments on metadata. In 2014, the CJEU declared the Data Retention Directive invalid in its *Digital Rights Ireland* judgment.[110] That directive obliged member states to adopt laws that require telecom and internet access providers to retain metadata of all their customers for a period of up to two years, for intelligence and law enforcement purposes.

The CJEU invalidated the directive because it interfered disproportionately with the EU Charter's privacy and data protection rights, and did not provide sufficient safeguards against arbitrariness and abuse. The CJEU noted that such metadata retention "is likely to generate in the minds of the persons concerned the feeling that their private lives are the subject of constant surveillance."[111]

After this judgment of the CJEU, many member states withdrew their national data retention laws.[112] However, some member states, including Sweden and the UK, kept national data retention laws in place. National courts in those two countries asked the CJEU whether such national data retention laws were allowed under EU law. In the *Tele 2/Watson* case,[113] the CJEU ruled, roughly summarized, that EU member states are not allowed to impose a blanket obligation on telecom and internet access providers to store metadata of all users of electronic communication services. According to the CJEU, such mass metadata retention, even if it aims to help catch criminals or terrorists, is disproportional, and therefore violates people's privacy and data protection rights.

---

108　P.G. and J.H. v. United Kingdom, 2001-IX Eur. Ct. H.R.
109　*See* Copland v. United Kingdom, 2007-I Eur. Ct. H.R.
110　Cases C-293/12 and C-594/12, Digital Rights Ireland and Seitlinger and Others v. Ireland, 2014 E.C.R. 54.
111　*See id.* at ¶ 37.
112　Mark Cole & Franziska Boehm, Data Retention After the Judgement of the Court of Justice of the European Union (June 30, 2014) (unpublished manuscript), http://hdl.handle.net/10993/17500.

The CJEU added that metadata enable "establishing a profile of the individuals concerned, information that is no less sensitive, having regard to the right to privacy, than the actual content of communications."[114] Moreover, mass metadata surveillance threatens the right to receive and impart information: the retention of metadata can "have an effect on the use of means of electronic communication and, consequently, on the exercise by the users thereof of their freedom of expression, guaranteed in Article 11 of the Charter."[115]

The CJEU confirmed this approach in the *Schrems* case, in which it invalidated the "Safe Harbor" decision that formed an important legal basis for the export of personal data to companies in the United States. According to the CJEU, the Commission's decision was invalid, because the Commission did not sufficiently investigate whether the Safe Harbor scheme offered a level of protection for personal data equivalent to the protection in the EU. This investigation was necessary, since it appeared that U.S. intelligence agencies had access to communications and metadata on a massive scale. According to the CJEU, mass surveillance of communications *content* violates the "essence" of the right to privacy.[116] If a law (such as the Data Retention Directive) infringes the essence of a right, that law is, by definition, illegal. In such a case, there is no need to assess whether the law constitutes a proportionate interference with a right. Hence, the CJEU makes a similar distinction between content and metadata as the ECtHR. Although metadata are protected under communications confidentiality, the CJEU sees metadata as normally less sensitive than content.

A similar distinction between content and metadata is made in the ePrivacy Directive, under which communication providers may only process metadata if they meet strict rules. In principle, such providers are only allowed to process metadata for transmission and billing purposes, and only as long as this is necessary. Additionally, providers may process metadata after prior consent of the individual, and only for a limited number of purposes.[117] But providers are in principle not allowed to touch communications *content* without the permission of the users. The 2017 ePrivacy proposal makes a similar distinction between content and metadata.

---

113  Joined Cases C-203/15 and C-698/15 Tele 2/Watson, ECLI:EU:C:2016:970.
114  *Tele 2/Watson*, ECLI:EU:C:2016:970, ¶ 99.
115  *Id*. at ¶ 101.
116  Case C-362/14, Maximillian Schrems v. Data Protection Commissioner, 2015 E.C.R. 117, ¶ 94.

However, the distinction between content and metadata is coming increasingly under pressure. First, it is becoming harder to distinguish content from metadata. Second, content and metadata can be equally sensitive and revealing. For example, it is contentious whether the subject line of an email message should be regarded as content or metadata. And regarding web browsing, URLs can be seen as content or as metadata. A URL, such as *<https://en.wikipedia.org/wiki/Theoretical_Inquiries_in_Law>,* can provide information about content. Third, metadata are easier to analyze than communications content. Fourth, collecting metadata (rather than content) enables parties to capture data about millions of people, because storing metadata is usually cheaper than storing content.

In conclusion, under the ECHR and under EU law, metadata are granted protection, but normally less than the communications content. This difference can partly be explained by the fact that communication service providers must process some metadata to provide their services. The law is still developing regarding the distinction between content and metadata, and how the law should deal with that distinction requires more research and debate.

### D. Protection after Transmission of Communications?

Does the right to communications confidentiality also apply after communications are transported or transmitted? As noted above, the right to communications confidentiality is historically connected to the state postal and telecommunications monopoly. This state monopoly made communications more vulnerable to interception during transport. Nowadays, communication has also become vulnerable outside of the strict transport phase. For example, communication is often stored on servers of communication service providers. Such developments raise the question whether communications confidentiality should also protect communication outside of the transport phase.

#### 1. Article 8 of the ECHR

Older case law took the position that the protection of "correspondence" under Article 8 of the ECHR ended after delivery of the letter.[118] The underlying assumption seemed to be that correspondence only needs extra protection during transport by the state postal company. Considering the historical background of the right to communications confidentiality, the assumption was defendable before the state monopoly was abolished.

---

117 *See* Directive 2002/58/EC, *supra* note 28, at arts. 5-6, 9.
118 *See, e.g.,* G., S. & M. v. Austria, App. No. 9614/81, 34 Eur. Comm'n H.R. Dec. & Rep. 119 (1983).

However, in 1992 the ECtHR extended the protection to communications after the transport phase. In the *Niemietz* case, the ECtHR applied article 8 for the first time to letters after delivery.[119] The case involved business correspondence that was obtained by the police during a search at the recipient's home. And in the 2013 *Bernh Larsen Holding* case, the ECtHR gave a similar broad protection to digital communications stored on a company server. In this case, the Norwegian tax authorities had ordered a company to provide a copy of all data stored on a shared server. The ECtHR said: "The imposition of that obligation on the applicant companies constituted an interference with their 'home' *and undoubtedly concerned their 'correspondence' and material that could properly be regarded as such* for the purposes of Article 8."[120]

The above case-law shows that Article 8 of the ECHR also protects communications after the transport has ended, regardless of the nature of the communication or the technology used.

2. EU law

It may be assumed that Article 7 of the EU Charter also protects communications after transport. After all, the EU Charter offers at least the same protection as the ECHR. However, under the ePrivacy Directive, the status of communications after transport is less clear. The main confidentiality provision of the ePrivacy Directive (Article 5(1)) does not make it explicit whether it protects communications after transport.

The ePrivacy Directive also aims to protect people's communication *devices*, in Article 5(3). Article 5(3) of the ePrivacy Directive protects information, and hence also communications, stored on a user's device. The ePrivacy Directive assumes that people's devices are part of their private sphere, and therefore deserve legal protection.[121]

Roughly summarized, under Article 5(3) of the ePrivacy Directive, parties can only store or access information on a user's device if that user has consented to it. Article 5(3) applies, for instance, when a company copies somebody's address book or emails from his or her phone.[122] The provision also applies to cookies, because cookies constitute information that is stored on, or read from, a device. There are exceptions to the consent requirement, for instance for cookies (etc.) that are necessary for transmission or for a service requested by the user.[123] Article 5(3) does not protect stored communications in the cloud.[124]

---

119  Niemietz v. Germany, 251-B Eur. Ct. H.R. (ser. A) (1992).
120  Bernh Larsen Holding AS and others v. Norway, CE:ECHR:2013:0314JUD002411708, ¶ 106 (2013). For a discussion, see Arnbak, *supra* note 11.
121  Directive 2002/58/EC, *supra* note 28, at recital 24. On the rationales for art. 5(3), see Frederik J. Zuiderveen Borgesius, *Personal Data Processing for Behavioural Targeting: Which Legal Basis?,* 5 Int'l Data Privacy L. 163 (2015).

Regarding communications after transmission, the legal situation under the 2017 ePrivacy proposal is similar to that under the ePrivacy Directive. It is unclear to what extent the main confidentiality provision (Article 5) of the ePrivacy proposal protects communication after transmission.[125] Article 8 of the ePrivacy proposal protects information (including communications) that is stored on a user's device — but does not protect communications stored in the cloud.

It is contentious to what extent the ePrivacy rules regarding communications confidentiality should protect communications after the transmission phase. European data protection authorities and others argue that communications should also be protected after transmission.[126] The Article 29 Working Party (consisting of the 28 EU data protection authorities) says that a restriction to the transmission phase "is based on a conceptual framework of communications which is outdated."[127] The Article 29 Working Party adds: "communication between subscribers of the same cloud-based services (for instance webmail providers) will often entail only very little conveyance: sending a mail would mostly involve reflecting this in the database of the provider, rather than actually sending communications between two parties."[128] Moreover, people that use communication services may not even realize whether their communications are in transit or stored.

---

122 On copying a contact list, see *Canadian and Dutch Data Privacy Guardians Release Findings from Investigation of Popular Mobile App*, Autoriteit Persoonsgegevens (Oct. 13, 2013), https://autoriteitpersoonsgegevens.nl/en/news/canadian-and-dutch-data-privacy-guardians-release-findings-investigation-popular-mobile-app.
123 Directive 2002/58/EC, *supra* note 28, at art. 5(3). *See also* Eleni Kosta, *Peeking into the Cookie Jar: The European Approach Towards the Regulation of Cookies*, 21 Int'l J.L. & Info. Tech. 380 (2013).
124 Article 5(3) speaks of "terminal equipment"; that concept excludes cloud storage. "Terminal equipment" is defined in Directive 2008/63/EC of the European Parliament and of the Council of 20 June 2008 on Competition in the Markets in Telecommunications Terminal Equipment, art. 5(3), 2008 O.J. (L 162).
125 For comments on art. 5, see Zuiderveen Borgesius et al., *supra* note 94, at 53-54.
126 *European Data Protection Supervisor Opinion*, *supra* note 4; *Opinion of the Article 29 Data Protection Working Party on the Proposed Regulation for the ePrivacy Regulation (2002/58/EC)*, 26, WP 247, 01/2017 (2017). *See also* Zuiderveen Borgesius et al., *supra* note 94, at 54-55.
127 *Article 29 Data Protection Working Party*, WP247, at 26.

But businesses, especially Silicon Valley companies that provide over the top services such as webmail, have lobbied for limiting the ePrivacy Regulation's scope of protection to the transmission phase. Such a limitation would put many over the top services outside of the scope of protection, because these services do not consist of conveying signals, but are delivered on top of communication networks (offered by, for instance, internet access providers). Many over the top service providers believe that the ePrivacy Regulation's rules regarding electronic communications are too strict.

We think that the ePrivacy Regulation should, at a minimum, protect communications when a service provider stores them in the cloud as part of the communication service. Many service providers, such as webmail providers, store people's communications by default after the transmission phase (in the narrow sense) has ended. This storage could be seen as a form of extended transmission. Hence, it makes sense to extend the protection until after the transmission phase, also because there are similar trust issues at stake as in the context of more traditional communication services.

We do not believe that the protection of the GDPR is sufficient for communications outside of the strict transmission phase, as some Silicon Valley companies seem to suggest. First, the GDPR only protects personal data relating to identifiable natural persons, which means that the GDPR would not protect all communications. For instance, some business-to-business communications may not be tied to personal data. We saw, however, that human rights case law on confidentiality of communications does require protection of business communications.

Second, the GDPR is written for all kinds of situations and contains open and relatively lenient norms, which are not tailored to protecting communications. For example, in many situations, the GDPR allows a company to process personal data (including communications that qualify as personal data) without the individual's consent. To illustrate: a company can process personal data if, in short, it has a legitimate interest to use the data, and that interest overrides the individual's interests.[129] In consideration of the human rights and trust issues at stake, it is more appropriate to have a regime (as in the ePrivacy Directive) that prohibits interference with communications, unless under narrowly defined specific circumstances, or when the individual has given prior consent.

---

128  *Id.* at 26.
129  GDPR, *supra* note 24, at art. 6(1)(f).

In sum, the right to communications confidentiality also applies after the transport or transmission of the communication has ended. The rules in the ePrivacy Regulation should also protect communications outside of the transmission phase.

## Conclusion

In this Article, we have discussed the right to communications confidentiality in Europe, to assess whether communications confidentiality requires extra protection, in separate rules, in addition to the GDPR.

We have seen that the right to communications confidentiality protects not only privacy-related interests, but also other interests which are more related to freedom of expression. Moreover, the right to communications confidentiality protects the trust that society as a whole has in communication services. Protecting that trust was important when the state had a postal and telecommunications monopoly. Nowadays, communication services are mostly offered by companies. But trust in communication services remains essential for communications, commercial transactions, e-government, and participation in democratic processes.

Because the right to communications confidentiality protects different values than the rights to privacy and to personal data protection, it has a different scope. The right protects not only private communications, but all communications, regardless of their nature and regardless of the technology used.

Some aspects of the right to communications confidentiality remain controversial. For instance, metadata can be just as sensitive as communications content — but how the law should protect metadata requires more research and debate. Another controversial topic is to what extent the right should offer protection after the transmission phase. Since trust in communication services is one of the core values protected by the right, it makes sense to protect communications and metadata in the cloud.

We conclude that separate protection of confidentiality of communications, next to the GDPR, is justified and necessary. An adequate protection of confidentiality of communications is crucial — not only for individuals, but for society as a whole.